\documentclass[aps,PRE]{revtex4-1}
\usepackage{graphicx}
\usepackage{float}

\newcommand{\kb}{k_{\mathrm{B}}}
\usepackage{amsmath}

\begin{document}
\title{
Non-linear Fluctuating Hydrodynamics with many conserved fields: the case of a 3D anharmonic chain
}

\author{R. Barreto$^{1}$}
\email{rbarreto@ungs.edu.ar}

\author{M. F. Carusela$^{1,2}$}
\email{flor@campus.ungs.edu.ar}

\author{A. G. Monastra$^{1,2}$}
\email{amonast@ungs.edu.ar}

\affiliation{$^1$ Instituto de Ciencias, Universidad Nacional de Gral. Sarmiento, Los Polvorines, Buenos Aires, Argentina}
\affiliation{$^2$ Consejo Nacional de Investigaciones Cient\'\i ficas y T\'ecnicas, Argentina}

\begin{abstract}
We propose a model for a chain vibrating in 3 dimensions (3D), with first neighbors anharmonic interatomic potential which depends  on their distance, and subjected to an external tension.

In the framework of the non-linear fluctuating hydrodynamic theory (NLFHT), that was successfully applied to 1D chains, we obtain a heat mode, two longitudinal and four transverse sound modes. We compute their spatio-temporal correlations comparing the theoretical results with molecular dynamics simulations, finding a good agreement for high temperatures. We find that the transverse sound modes behave diffusively, meanwhile the heat and longitudinal sound modes superdiffusively, exploring their possible scaling functions and characteristic exponents.
\end{abstract}


\maketitle

\section{Introduction}

Anomalous transport have been observed in a wide variety of systems revealing that this phenomenon is ubiquitous in nature.
Its emergence can be found in processes taking place in plasmas, glassy materials, porous media, polymers and flexible filaments in viscoelastic media, biological cells, heat conduction, chemical reaction-diffusion, epidemic spreading, just to mention some \cite{libro1,libro2,cells}.

Low dimensional systems, such as polymers, nanowires, and nanoribbons, generally present anomalous thermal transport properties, in other words Fourier's Law is not fulfilled. This anomalous behaviour have been studied theoretically, numerically and also experimentally \cite{dhar2,li}.

Among different theoretical approaches to study anomalous thermal transport, it has been recently proposed a nonlinear extension of fluctuating hydrodynamics. It is a solid scheme that can be applied to general dynamics as long as the interactions are local and translationally invariant, and there are locally conserved fields. Under these conditions it is possible to compute steady state average currents, that are functions of the  conserved fields \cite{Spohn2014}.

 Models for a chain of atoms with anharmonic interactions  have been considered to study thermal transport under this approach. However, in these models the atoms were allowed to vibrate only in one dimension and periodic boundary conditions were usually considered \cite{EvenOlla2014,Spohn2014,sponh2,dhar,Spohn2015}.  In more realistic systems the interaction depends on the relative position between atoms, so vibrations in different directions are coupled. Moreover, experimental conditions usually require setups that are suspended or clamped by their ends and subject to external stresses.
 
 In this context, we extend recent works on anomalous thermal transport in low dimensional systems \cite{nos1,nos2} in the framework of the NLFHT to the case of stressed anharmonic chains with 3D motion. The theoretical results are checked by molecular dynamics simulations. 
 
 In section II, we present the model and the equations of motion, calculating average values in the canonical ensemble. In Section III, we apply NLFHT to our model transforming the elongation, momentum and energy (fields) of the particles to a normal mode basis, characterizing their correlations. We also discuss the scaling functions in the framework of mode-coupling theory (MCT).
 In section IV we compute through molecular dynamics simulations the evolution and spatio-temporal correlations of the fields, comparing with the theoretical predictions. Finally, in Section V we present some brief conclusions.

\section{Model and global equilibrium}

The model consists of a chain of $N+1$ identical particles of mass $m$, labeled with index $0 \leq n \leq N$. The first particle $(n = 0)$ is fixed in the coordinate origin. The other $N$ particles can move in the three $x$-$y$-$z$ directions and are subjected to a nearest neighbor potential $V(r)$, that depends only on the distance between them. At this point, we consider a general potential with the condition to have one minimum at a finite equilibrium distance $r_0$. Moreover, the last particle $n=N$ is subjected to an external constant force ${\bf F}$, which provides a tension along the chain. 

\begin{figure}
\begin{center}
\includegraphics[width=20pc]{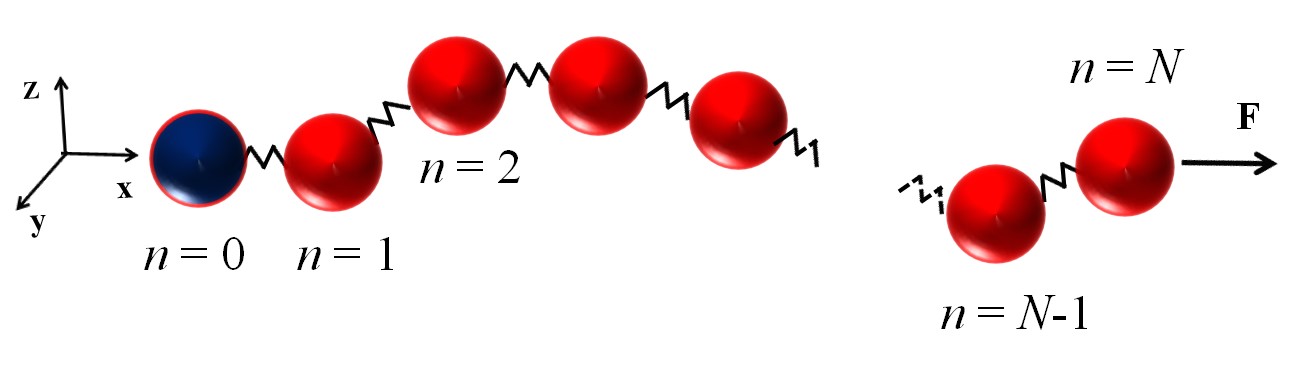}
\end{center}
\caption{\label{modelo} (Color online) Sketch of the chain system. The blue particle ($n =0$) is fixed. The other particles can move in the three directions.}
\end{figure}

In terms of positions ${\bf R}_n$ and momenta ${\bf P}_n$ of the particles, the Hamiltonian is

\begin{gather}
{\cal H} (\{ {\bf R}_n, {\bf P}_n \})= \nonumber \\
\frac{1}{2 m} \sum_{n = 1}^{N} | {\bf P}_n |^2 + \sum_{n = 1}^{N} V(| {\bf R}_n - {\bf R}_{n - 1}|) - {\bf F} \cdot {\bf R}_N \ .
\end{gather}
Due to the nearest neighbor interactions, the equations of motion given by this Hamiltonian can be rewritten in a simpler way in terms of the inter-particle coordinates ${\bf r}_n = {\bf R}_n - {\bf R}_{n - 1}$, and the force that particle $(n-1)$ does on particle $n$

\begin{equation}
{\bf f}_n = - V'(r_n) \frac{{\bf r}_n}{r_n} \ ,
\end{equation}
with $r_n = | {\bf r}_n |$, finally giving 

\begin{eqnarray}
\frac{\textnormal{d} {\bf r}_n }{ \textnormal{d} t } &=& \frac{ {\bf P}_n }{m} - \frac{ {\bf P}_{n - 1} }{m} \ , \label{EqMotionElongation} \\
\frac{\textnormal{d} {\bf P}_n }{ \textnormal{d} t } &=& {\bf f}_n - {\bf f}_{n+1} \ . \label{EqMotionMomentum}
\end{eqnarray}
These equations are valid for $1 \leq n \leq N$, reminding that ${\bf P}_0 = 0$, and defining ${\bf f}_{N+1} = - {\bf F}$.

The energy per site is defined as
\begin{equation} 
\epsilon_n = \frac{1}{2 m} |{\bf P}_n|^2 + V(r_n) \ .
\end{equation}
This is the kinetic energy of particle $n$ plus the potential energy between particle $(n-1)$ and $n$. We stress that this is an arbitrary definition, because the potential energy of a bond is shared by two particles. The derivative with respect to time of this site energy is

\begin{equation} \label{EqMotionEnergy}
\frac{\textnormal{d} \epsilon_n }{ \textnormal{d} t } = \frac{1}{m} {\bf f}_n \cdot {\bf P}_{n-1} - \frac{1}{m} {\bf f}_{n+1} \cdot {\bf P}_n \ ,
\end{equation}
which corresponds to a local conservation of energy. In the limit $N \rightarrow \infty$, equilibrium statistics of the system tend to the same results both in the microcanonical or canonical ensembles.
In the canonical ensemble, the probability density to find the system in a particular configuration is ${\cal P} (\{ {\bf R}_n , {\bf P}_n \}) \textnormal{d} \Gamma = Z^{-1} \exp ( - \beta {\cal H}) \textnormal{d} \Gamma$, with $\beta = 1/(\kb T)$ the inverse temperature, $Z$ the canonical partition function, and $\textnormal{d} \Gamma = \prod_n \mathrm{d}{\bf R}_n  \mathrm{d} {\bf P}_n $ the differential of phase space. The average value of any quantity is computed integrating it over the phase space with this probability density. Although the canonical variables are ${\bf R}_n$ and ${\bf P}_n$, we can again take advantage of the structure of the Hamiltonian, and change the space coordinates from ${\bf R}_n$ to the ${\bf r}_n$. This transformation has a Jacobian equals to one. Moreover, ${\bf R}_N = \sum_{n=1}^N {\bf r}_n$, and the probability density factorizes completely to

\begin{gather}
{\cal P} (\{ {\bf r}_n , {\bf P}_n \}) \textnormal{d} \Gamma = \nonumber \\ \frac{1}{Z} \prod_{n = 1}^{N} \exp \left( - \frac{\beta}{2 m} |{\bf P}_n|^2 \right) \mathrm{d} {\bf P}_n \exp \left( - \beta V (r_n) + \beta {\bf F} \cdot {\bf r}_n \right) \mathrm{d}{\bf r}_n \ .
\end{gather}
We see explicitly that each component of momentum is a random Gaussian variable. The elongations ${\bf r}_n$ for different bonds are also decorrelated, although the three cartesian components of each one are correlated through the term $V (r_n)$. Integrating this probability density we obtain the partition function

\begin{equation}
Z(\beta, {\bf F}) = \left( \frac{2 \pi m}{ \beta} \right)^{3 N/2}  \zeta (\beta, {\bf F})^N \  ,
\end{equation}
with $\zeta (\beta, {\bf F}) = \int \exp \left[ - \beta (V (r) -  {\bf F} \cdot {\bf r} ) \right] \mathrm{d}{\bf r}$. To compute this function and other mean values on the canonical ensemble, it is useful to change to spherical coordinates, putting the polar axis in the direction of the external force. In this case the integrals in the angular coordinates are easily performed. To simplify the explicit expressions it is also useful to define the following radial integrals

\begin{eqnarray}
S_k (\beta, F) &=& \int_0^{\infty} \mathrm{d}r  \  r^k \sinh (\beta F r) \exp \left[ - \beta V (r) \right]  \ , \\
C_k (\beta, F) &=& \int_0^{\infty} \mathrm{d}r  \  r^k \cosh (\beta F r) \exp \left[ - \beta V (r) \right] \ .
\end{eqnarray}
These integrals converge only for potentials growing faster than linear, i.e.  $V(r)/r \rightarrow \infty$ for $r \rightarrow \infty$. This condition corresponds to interparticle forces at large elongations being stronger than the applied external force. 
In terms of these radial integrals, the partition function, the mean values of the elongation, and the energy are expressed as 

\begin{eqnarray}
\zeta &=& \frac{4 \pi}{\beta F} S_1 \ ,\\
\langle x \rangle &=& \frac{C_2}{S_1} - \frac{1}{\beta F} = \ell (\beta, F) \ , \\
\langle \epsilon \rangle &=& \frac{3}{2 \beta} + \langle V \rangle = e(\beta, F) \ .
\end{eqnarray}
The functions $\ell$ and $e$ are of relevance for the following theoretical calculations. To pass to the microcanonical ensemble, in the limit $N \rightarrow \infty$, one can invert the relations to obtain $\beta(\ell,e)$ and $F(\ell,e)$.

The mean values of momentum and force are $\langle P_x \rangle =  \langle P_y \rangle = \langle P_z \rangle = 0$, $\langle f_x \rangle = -F$, and $\langle f_y \rangle = \langle f_z \rangle = 0$. It is also possible to compute the second moments, obtaining

\begin{eqnarray}
\langle x^2 \rangle &=& \frac{S_3}{S_1} - \frac{2}{\beta F} \frac{C_2}{S_1} + \frac{2}{(\beta F)^2} \ ,\\
\langle y^2 \rangle &=& \langle z^2 \rangle = \frac{1}{\beta F} \frac{C_2}{S_1} - \frac{1}{(\beta F)^2} = \frac{\ell}{\beta F} \ , \\
\langle P^2_x \rangle &=& \langle P^2_y \rangle = \langle P^2_z \rangle = \frac{m}{\beta} \ ,\\
\langle \epsilon^2 \rangle &=& \frac{15}{4 \beta^2} + \frac{3}{\beta} \langle V \rangle + \langle V^2 \rangle \ .
\end{eqnarray}

We can generalize this ensemble to a situation where there is a center of mass velocity ${\bf V}_0$ (it corresponds to a constant velocity of the first particle $n=0$ originally fixed), and an arbitrary direction of the force. With these changes, the mean values become  

\begin{eqnarray}
&\langle x \rangle = \ell (\beta, F) \frac{F_x}{F} \ , \hspace{0.5cm} \langle y \rangle = \ell (\beta, F) \frac{F_y}{F} \ ,  \nonumber \\
&\langle z \rangle = \ell (\beta, F) \frac{F_z}{F}\ ,  \nonumber \\ \nonumber\\
&\langle P_x \rangle = m V_{0x} \ , \hspace{0.5cm}
\langle P_y \rangle = m V_{0y} \ , \nonumber \\
&\langle P_z \rangle = m V_{0z} \ ,\nonumber \\ \nonumber\\
&\langle \epsilon \rangle = \frac{1}{2} m |{\bf V}_0|^2 + e(\beta, F) \ .
\end{eqnarray}
where $\ell$ and $e$ are the same previously defined functions.

\section{Nonlinear fluctuating hydrodynamic theory}

\subsection{Mesoscopic fields and currents}

Following the theoretical developments in \cite{Spohn2014} applied to anharmonic chains, we extend it to our model for a stressed chain vibrating in 3D. We first define seven microscopic fields

\begin{gather}
g_1 (n, t) = x_n (t) \ , \hspace{0.25cm} g_2 (n, t) = y_n (t) \ , \hspace{0.25cm} g_3 (n, t) = z_n (t) \ , \nonumber \\
g_4 (n, t) = P_{x, n} (t) \ , \hspace{0.18cm}  
g_5 (n, t) = P_{y, n} (t)  \ , \hspace{0.18cm}  
g_6 (n, t) = P_{z, n} (t) \ , \nonumber \\
g_7 (n, t) = \epsilon_n (t) \ , \nonumber
\label{MicroFields}
\end{gather}
and their corresponding microscopic currents
\begin{gather}
J_1 (n, t) = -\frac{1}{m} P_{x, n} (t) \ , \hspace{0.25cm}
J_2 (n, t) = -\frac{1}{m} P_{y, n} (t) \ , \nonumber \\
J_3 (n, t) = -\frac{1}{m} P_{z, n} (t) \ , \hspace{0.25cm}
J_4 (n, t) = f_{x, n+1} (t) \ , \nonumber \\
J_5 (n, t) = f_{y, n+1} (t) \ , \hspace{0.25cm}
J_6 (n, t) = f_{z, n+1} (t) \ , \nonumber \\
J_7 (n, t) = \frac{1}{m} {\bf f}_{n+1} (t) \cdot {\bf P}_{n} (t) \ . \label{MicroCurrents}
\end{gather}
By these definitions, the equations of motion (\ref{EqMotionElongation}) and (\ref{EqMotionMomentum}) and conservation of energy (\ref{EqMotionEnergy}) can be written in the compact form

\begin{equation}
\frac{\textnormal{d}}{\textnormal{d} t} g_{\alpha} (n, t) + J_{\alpha} (n,t) - J_{\alpha} (n-1,t) = 0 \ ,
\end{equation}
which are the Euler equations. The index $n$ is discrete, and the equations are valid for $n=1$ up to $n=(N-1)$, being the first particle ($n=0$) fixed, and the last particle ($n =N$) is subjected to the external force ${\bf F}$.

To connect these fields to the results of the generalized canonical ensemble, we define for any arbitrary microscopic quantity $h_n (t)$ a coarse-graining or mesoscopic average

\begin{equation}
\langle h \rangle (n, t) =  \sum_{n'=1}^N h_n \Theta (n' - n) \ ,
\end{equation}
with $\Theta (x)$ a smoothing function with the following properties: positive $\Theta (x) \geq 0 \ \forall x$, normalized $\int \Theta (x) \textnormal{d} x = 1$, and with a finite variance $\sigma^2 = \int x^2 \Theta (x) \textnormal{d} x$ such that $1 \ll \sigma \ll N$, typically a Gaussian function. Now $n$ becomes a continuous variable along the chain, instead of a discrete index. This mesoscopic average is valid far from the ends of the chain $( 3 \sigma \lesssim n \lesssim N - 3 \sigma)$.

This coarse-graining, in a statistical sense, is related to a local thermal equilibrium of the chain around $n$. The averaged fields and currents become smooth functions, not varying too much in the scale of few sites, finally providing the (continuous) Euler equations

\begin{equation}
\frac{\partial}{\partial t} {\cal G}_{\alpha} (n, t) + \frac{\partial}{\partial n} {\cal J}_{\alpha} (n,t) = 0 \ ,
\end{equation}
where ${\cal G}_{\alpha} = \langle g_{\alpha} \rangle$ and ${\cal J}_{\alpha} = \langle J_{\alpha} \rangle$ are the smoothed fields and currents, respectively. A local microcanonical equilibrium can be mapped to the general canonical ensemble defined by the seven parameters ${\bf V}_0, {\bf F}$ and $\beta$ (local center of mass velocity, local tension and local temperature), which can smoothly vary over the chain, not far from the global equilibrium, giving for the smooth fields

\begin{eqnarray}
{\cal G}_1 &=& \ell \frac{F_x}{F} \ , \hspace{0.65cm}
{\cal G}_2 = \ell \frac{F_y}{F} \ , \hspace{0.65cm}
{\cal G}_3 = \ell \frac{F_z}{F} \ , \nonumber \\
{\cal G}_4 &=& m V_{0x} \ , \hspace{0.5cm}
{\cal G}_5 = m V_{0y} \ , \hspace{0.5cm}
{\cal G}_6 = m V_{0z} \ , \nonumber \\
{\cal G}_7 &=& \frac{1}{2} m |{\bf V}_0|^2 + e \ .
\end{eqnarray}
One can invert these relations to obtain the seven parameters of the ensemble in terms of the fields
\begin{eqnarray}
F_x &=&  \frac{F}{\ell} {\cal G}_1 \ , \hspace{0.5cm}
F_y =  \frac{F}{\ell} {\cal G}_2 \ , \hspace{0.5cm}
F_z =  \frac{F}{\ell} {\cal G}_3 \ , \nonumber \\
V_{0x} &=& \frac{1}{m} {\cal G}_4 \ , \hspace{0.5cm}
V_{0y} = \frac{1}{m} {\cal G}_5 \ , \hspace{0.5cm}
V_{0z} = \frac{1}{m} {\cal G}_6 \ , \nonumber \\
e &=& {\cal G}_7 - \frac{1}{2 m} ({\cal G}_4^2 + {\cal G}_5^2 + {\cal G}_6^2) \ ,
\end{eqnarray}
where now $\ell = \sqrt{{\cal G}_1^2 + {\cal G}_2^2 + {\cal G}_3^2}$ is also a function of the smooth fields. Temperature $\beta$ and force modulus $F$ are defined implicitly in terms of $\ell$ and $e$. This inversion allows to express the local mesoscopic currents in terms of the averaged fields:

\begin{eqnarray}
{\cal J}_1 &=& -\frac{1}{m} {\cal G}_4 \ , \hspace{0.5cm}
{\cal J}_2 = -\frac{1}{m} {\cal G}_5 \ , \hspace{0.5cm}
{\cal J}_3 = -\frac{1}{m} {\cal G}_6 \ , \nonumber \\
{\cal J}_4 &=& -\frac{F}{\ell} {\cal G}_1 \ , \hspace{0.5cm}
{\cal J}_5 = -\frac{F}{\ell} {\cal G}_2 \ , \hspace{0.5cm}
{\cal J}_6 = -\frac{F}{\ell} {\cal G}_3 \ , \nonumber \\
{\cal J}_7 &=& -\frac{1}{m} \frac{F}{\ell} ({\cal G}_1 {\cal G}_4 + {\cal G}_2 {\cal G}_5 + {\cal G}_3 {\cal G}_6) \ . \label{CurrentsInFields} 
\end{eqnarray}
For the last current ${\cal J}_7$, we have used that the microscopic force, which only depends on spatial coordinates, is decorralated from the microscopic momentum, as in the general canonical ensemble.

We see explicitly that the currents are non-linear functions of the fields (except for the first three components). Considering that the local equilibrium is not far from a global equilibrium, the currents can be expanded up to second order.

The global equilibrium is a uniform state over the chain, where the fields are
\begin{equation}
\vec{\cal G}_0 = (\ell_0, 0, 0, 0, 0, 0, e_0) \ .
\end{equation}
This state can be defined by a global temperature $\beta_0$ and external tension $F_0$ in the $x$ direction, therefore $\ell_0 = \ell (\beta_0, F_0)$ and $e_0 = e (\beta_0, F_0)$. On the other hand, we can constrain the last particle to move in a plane $y$-$z$ at distance $L_0$ from the first particle, with a fixed total energy of the chain $E_0$, then $\ell_0 = L_0/N$ and $e_0 = E_0/N$. In the thermodynamic limit $N \rightarrow \infty$ both conditions would give the same results.

Around this global equilibrium, we expand the currents up to second order

\begin{equation}
\begin{split}
{\cal J}_{\alpha} ( \vec{\cal G} ) = {\cal J}_{\alpha} ( \vec{\cal G}_0 ) + \sum_{\beta = 1}^7 A_{\alpha \beta} u_{\beta} + \nonumber \\ \frac{1}{2} \sum_{\beta = 1}^7 \sum_{\gamma = 1}^7 H^{\alpha}_{\beta \gamma} u_{\beta} u_{\gamma} + {\cal O} (u^3) \ , 
\end{split}
\end{equation}
with the Jacobian and Hessians matrices

\begin{equation}
A_{\alpha \beta} = \frac{ \partial {\cal J}_{\alpha} }{ \partial {\cal G}_{\beta} } ( \vec{\cal G}_0 ) \ ,
\end{equation}

\begin{equation}
H^{\alpha}_{\beta \gamma} = \frac{ \partial^2 {\cal J}_{\alpha} }{ \partial {\cal G}_{\beta} \partial {\cal G}_{\gamma} } ( \vec{\cal G}_0 ) \ ,
\end{equation}
and $u_{\alpha} (n, t) = {\cal G}_{\alpha} (n, t) - {\cal G}_{0 \alpha}$, the fluctuations of the fields around the global equilibrium, which are typically small.

The matrix $A$ can be computed explicitly from Eqs. (\ref{CurrentsInFields}), in terms of the derivatives $F_{\ell} = \partial F / \partial \ell$, and $F_{e} = \partial F / \partial e$, giving

\begin{equation}
A_{\alpha \beta} = \left(
\begin{array}{ccccccc}
0 & 0 & 0 & -\frac{1}{m} & 0 & 0 & 0 \\
0 & 0 & 0 & 0 & -\frac{1}{m} & 0 & 0 \\
0 & 0 & 0 & 0 & 0 & -\frac{1}{m} & 0 \\
- F_{\ell} & 0 & 0 & 0 & 0 & 0 & - F_e \\
0 & -\frac{F}{\ell} & 0 & 0 & 0 & 0 & 0 \\
0 & 0 & -\frac{F}{\ell} & 0 & 0 & 0 & 0 \\
0 & 0 & 0 & -\frac{F}{m} & 0 & 0 & 0 
\end{array} 
\right) \ .
\end{equation}
Although the force $F$ is an implicit function of $\ell$ and $e$, its derivatives can be computed in terms of derivatives of $\ell$ and $e$ with respect to $\beta$ and $F$ \cite{Spohn2014}, which in turn can be expressed in terms of the second moments of the fields. In the same way, the Hessian matrices $H^{\alpha}$ are written in terms of first and second derivatives of the external force, needing up to third moments of the fields. Their full explicit expressions are in the Appendix.

We are interested in the spatio-temporal correlations between the fluctuation of the fields 
\begin{equation}
{C}_{\alpha, \beta} (n, t)= \left\langle u_{\alpha} (n_0, t_0)  u_{\beta} (n_0 + n, t_0 + t) \right\rangle   \ .
\end{equation}
This average is done over different reference sites $n_0$ far from the borders, and for different reference times $t_0$. For $n=0$ and $t=0$, this auto-correlation matrix can be computed explicitly from the second moments computed in the canonical ensemble:

\begin{equation}
C_{\alpha \beta}(0,0) =
\left(
\begin{array}{ccccccc}
C_{1 1} & 0 & 0 & 0 & 0 & 0 & C_{1 7} \\
0 & \frac{\ell}{\beta F} & 0 & 0 & 0 & 0 & 0 \\
0 & 0 & \frac{\ell}{\beta F} & 0 & 0 & 0 & 0 \\
0 & 0 & 0 & \frac{m}{\beta} & 0 & 0 & 0 \\
0 & 0 & 0 & 0 & \frac{m}{\beta} & 0 & 0 \\
0 & 0 & 0 & 0 & 0 & \frac{m}{\beta} & 0 \\
C_{1 7} & 0 & 0 & 0 & 0 & 0 & C_{7 7}
\end{array} 
\right) \ ,
\end{equation}
with 
\begin{eqnarray}
C_{1 1} &=& \langle x^2 \rangle - \langle x \rangle^2 = \langle \delta x^2 \rangle  \ ,\\
C_{1 7} &=& \langle x \epsilon \rangle - \langle x \rangle \langle  \epsilon \rangle = \langle \delta x \ \delta V \rangle \ , \\
C_{7 7} &=& \langle \epsilon^2 \rangle - \langle \epsilon \rangle^2 =  \frac{3}{2 \beta^2} + \langle \delta V^2 \rangle  \ ,
\end{eqnarray}
where for simplicity $\delta x = x - \langle x \rangle$, and $\delta V = V - \langle V \rangle$.

With the previous expansion of the currents, the Euler equations up to second order in the fields are

\begin{equation} \label{EulerEqsSmooth}
\frac{\partial u_{\alpha}}{\partial t} + \frac{\partial}{\partial n} \left(  \sum_{\beta = 1}^7 A_{\alpha \beta} u_{\beta} + \frac{1}{2} \sum_{\beta = 1}^7 \sum_{\gamma = 1}^7 H^{\alpha}_{\beta \gamma} u_{\beta} u_{\gamma} \right) = 0 \ .
\end{equation}

In these equations all fields are coupled. Nevertheless, up to linear order, the equations can be decoupled by diagonalizing the matrix $A$, which is guaranteed by the property $C A= C A^{\text{T}}$, being $C = C_{\alpha \beta}(0,0)$, which was checked explicitly in our model. This provides seven normal modes with eigenvalues

\begin{equation}
c_i = \{ 0, +c_x, -c_x, +c_y, -c_y, +c_z, -c_z \}  \ , 
\end{equation}
where
\begin{equation}
c_x = \frac{1}{\sqrt{m}} \sqrt{ F_{\ell} +F  F_{e} }  \ ,
\end{equation}
and
\begin{equation}
c_y = c_z = c_{\perp} = \sqrt{ \frac{F}{m \ell} }  \ .
\end{equation}
The solutions of the Euler equations up to linear order are travelling waves with velocities $c_i$. The mode with eigenvalue zero is called the heat mode. There are two longitudinal sound modes (each travelling in opposite directions along the chain), and four transverse sound modes, with degenerate sound velocities $c_{\perp}$. Explicit expressions of the eigenvectors are in the Appendix. In this mode representation, the original $u_{\alpha}$ fields are transformed to
\begin{equation}
\phi_{\alpha} (n, t) = \sum_{\beta = 1}^7 R_{\alpha \beta} u_{\beta} (n, t)  \ ,
\label{matrizR}
\end{equation}
where $R$ is the basis change matrix composed by the left eigenvectors of $A$ as rows. We order the modes in the following way:

\begin{table}[h]
    \centering
    \begin{tabular}{||c| c| c||}
        \hline
          
          $\phi_1$ & Heat mode &  \\
          $\phi_2$ & Longitudinal mode & $x+$\\
          $\phi_3$ & Longitudinal mode & $x-$\\
          $\phi_4$ & Transverse mode & $y+$\\
          $\phi_5$ & Transverse mode & $y-$\\
          $\phi_6$ & Transverse mode & $z+$\\
          $\phi_7$ & Transverse mode & $z-$\\
         \hline
    \end{tabular}
\end{table}
In this new basis the spatio-temporal correlations are:
\begin{equation}
{\cal S}_{\alpha, \beta} (n, t)= \left\langle \phi_{\alpha} (n_0, t_0)  \phi_{\beta} (n_0 + n, t_0 + t) \right\rangle  \ ,
\label{correlaciones}
\end{equation}
where the average is computed in the same way as it was explained for the original $u_{\alpha}$ fields.

\subsection{Noise, diffusion and spatio-temporal correlations}

After considering the coarse-graining smoothing of the elongation, momentum and energy fields, and their corresponding currents, it is necessary to analyze the short range fluctuations to characterize the broadeaning of correlations in time.

On top of the smooth fluctuations along the chain, there are short range fluctuations that can be considered random. One can model these fluctuations adding diffusion and noise to Eq. (\ref{EulerEqsSmooth})

\begin{equation}
\begin{split}
\frac{\partial u_{\alpha}}{\partial t} +\frac{\partial}{\partial n} \Big(  \sum_{\beta = 1}^7 A_{\alpha \beta} u_{\beta} + \frac{1}{2} \sum_{\beta = 1}^7 \sum_{\gamma = 1}^7 H^{\alpha}_{\beta \gamma} u_{\beta} u_{\gamma} - \\
\frac{\partial}{\partial n} \sum_{\beta = 1}^7 D_{\alpha \beta} u_{\beta} + \sum_{\beta = 1}^7 B_{\alpha \beta} \xi_{\beta} \Big) = 0 \ , 
\label{NoisyEulerEqs}
\end{split}
\end{equation}
with $u_{\alpha}$ indicating the fields subjected to the stochastic fluctuations $\xi_{\beta} (n, t)$ which are random white noise components whose correlations are
\begin{equation} \label{NoiseCorrelations}
\langle \xi_{\alpha} (n, t) \xi_{\beta} (n', t') \rangle = \delta_{\alpha \beta} \delta (n- n') \delta (t -t') \ .
\end{equation}
$B_{\alpha \beta}$ is the noise strength matrix, which is diagonal \cite{sponh2}. The noise does not affect the first three elongation components, and due to the symmetry of the model in the transverse directions the matrix $B$ has the following structure

\begin{equation}
B_{\alpha \beta} = 
\left(
\begin{array}{ccccccc}
0 & 0 & 0 & 0 & 0 & 0 & 0 \\
0 & 0 & 0 & 0 & 0 & 0 & 0 \\
0 & 0 & 0 & 0 & 0 & 0 & 0 \\
0 & 0 & 0 & \sigma_{p_x} & 0 & 0 & 0 \\
0 & 0 & 0 & 0 & \sigma_{p_{\perp}} & 0 & 0 \\
0 & 0 & 0 & 0 & 0 & \sigma_{p_{\perp}} & 0 \\
0 & 0 & 0 & 0 & 0 & 0 & \sigma_{e}
\end{array} 
\right) \ .
\end{equation}
The symmetric diffusion matrix $D_{\alpha \beta}$ is related to the noise matrix and the correlations by the generalized fluctuation-dissipation theorem
\begin{equation} \label{DiffusionNoiseRelation}
D C + C D^{\text T} = B B^{\text T} \ .
\end{equation}
From this relation we obtain the diffusion matrix explicit and uniquely (see Appendix for its full expression).

In the mode basis, Eq. (\ref{NoisyEulerEqs}) transforms to:

\begin{equation}
\begin{split}
\frac{\partial \phi_{\alpha}}{\partial t} + \frac{\partial}{\partial n} \Big( c_{\alpha} \phi_{\alpha} + \sum_{\beta = 1}^7 \sum_{\gamma = 1}^7 G^{\alpha}_{\beta \gamma} \phi_{\beta} \phi_{\gamma} - \\ \frac{\partial}{\partial n} \sum_{\beta = 1}^7 \tilde{D}_{\alpha \beta} \phi_{\beta} + \sum_{\beta = 1}^7 \tilde{B}_{\alpha \beta} \xi_{\beta} \Big) = 0 \ ,
\end{split}
\label{NoisyEulerEqsModes}
\end{equation}
where  $\tilde{D}=R D R^{-1}$ and $\tilde{B}=R B$. In the new basis the relation Eq. (\ref{DiffusionNoiseRelation}) transforms to $\tilde{D}+\tilde{D}^{\text T}= \tilde{B} \tilde{B}^{\text T}$. The matrices $G^{\alpha}$ are obtained form the $H^{\alpha}$ matrices as:
\begin{equation}
G^{\alpha}=\frac{1}{2} \sum_{\beta=1}^{7} R_{\alpha \beta} (R^{-1})^{\text T}
H^{\beta} R^{-1} \ .
\end{equation}

As the modes travel with their own sound velocities, the crossed terms for the correlations ${\cal S}_{\alpha, \beta} (n, t)$ for $\alpha \neq \beta$ tends to zero for long times. In our model the only exceptions could be ${\cal S}_{4,6}$ and ${\cal S}_{5,7}$, because the sound velocities in the directions $y$ and $z$ are degenerated.

Given also the symmetry between modes $\phi_2$, $\phi_4$, and $\phi_6$ (travelling along the chain in positive direction) with modes $\phi_3$, $\phi_5$, and $\phi_7$ (travelling in the negative direction), the relevant correlations to study are ${\cal S}_{1,1}$, ${\cal S}_{2,2}$ and ${\cal S}_{4,4}$. 

For long time it is expected the following general scalings for the correlations:

\begin{eqnarray}
{\cal S}_{1, 1} (n, t) &=& t^{-\delta_1} f_1(n t^{-\delta_1}) \ ,\\
{\cal S}_{2, 2} (n, t)&=&t^{-\delta_2} f_2((n-c_x t)  t^{-\delta_2}) \ ,\\
{\cal S}_{4, 4} (n, t)&=&t^{-\delta_4} f_4((n-c_{\perp} t)  t^{-\delta_4}) \ .
\label{exponentes}
\end{eqnarray}

The longitudinal sound mode has a self-coupling quadratic term $G^{2}_{2 2}$ that does not vanish. Therefore, the corresponding Eq. (\ref{NoisyEulerEqs}) has the structure of a noisy Burgers equation whose solution is the scaling Kardar-Parisi-Zhang (KPZ) function, with a characteristic exponent $\delta_2 = 2/3$.

For the heat and transverse sound modes, the self-coupling terms $G^{1}_{1 1}$ and $G^{4}_{4 4}$ vanish. Therefore, to compute their auto-correlations ${\cal S}_{1, 1}$ and ${\cal S}_{4, 4}$, respectively, it is necessary to take into account subleading corrections. This can be done within the mode-coupling theory. In this approach the dynamical evolution of the correlations is given by integro-differential equations involving the following memory kernel \cite{Spohn2014}:
\begin{equation}
M_{\alpha \alpha}(n,t)=2 \sum_{\beta \gamma}(G^{\alpha}_{\beta \gamma})^{2} {\cal S}_{\beta \beta}(n,t) {\cal S}_{\gamma \gamma} (n,t)\ .
\label{kernel}
\end{equation}
For long times, due to the spread of the correlations at different velocities, the leading contributions to this kernel come from the diagonal terms, plus eventual contributions from the degenerate transverse modes.

For the transverse sound mode, due to the symmetries of the $G^{4}$ matrix, the kernel in Eq. (\ref{kernel}) vanishes. Under these conditions the equation for this mode reduces to a normal diffusion equation, with a Gaussian solution with a characteristic exponent $\delta_4=1/2$.

For the heat mode, due to the structure of the matrix $G^{1}$, the kernel has contributions from all the other six sound modes. If one considers only the coupling with the two longitudinal sound modes, whose correlation function approaches to a KPZ with exponent 2/3, one expects for the heat mode correlation a Levy function with exponent $\delta_1 = 3/5$. On the other hand, considering only a coupling with the four transverse sound modes, whose correlations are Gaussian functions, one would expect a Levy function with exponent $\delta_1 = 2/3$ for the heat mode correlation \cite{Spohn2015}.

Our model is more complex because it mixes both situations, so the exponent $\delta_1$ depends on the particular parameters, through the $G^1$ matrix elements. Nevertheless, the previous analysis indicates a superdiffusive behavior $\delta_1 > 1/2$.

\section{Molecular Dynamics simulations}

We apply the previous theoretical results to the case of a Fermi-Pasta-Ulam potential 

\begin{equation} \label{VFPU}
V (r) =  \frac{1}{2} k_2 (r - r_0 )^2 + \frac{1}{3} k_3 (r - r_0 )^3 + \frac{1}{4} k_4 (r - r_0 )^4 \ ,
\end{equation}
which can be considered as a general expansion of an arbitrary potential around its minimum. The quartic term $k_4$ should be bigger than zero, to guarantee the convergence of all integrals needed for the canonical averages.
We consider $k_2 = 1$, $k_3 = -5$ and $k_4 = 16$. These values guarantee the existence of only one minimum at $r=r_0$. On the other hand, $k_3$ tunes the asymmetry of the potential which  is negative in our case, to model a repulsive force at short distances.

We take $r_0$ as unit of length and $k_2 r_0^2$ as unit of energy and temperature.

Depending on temperature and on the applied tension the chain presents two different configurations. When the mean elongation $\ell$ is less than one, the chain is soft and the system has a huge number of available configurations, making difficult to explore numerically the full phase space. On the other hand, if $\ell > 1$, the chain is more tight and the available phase space for the system is drastically reduced.

\begin{figure*}
\centering
\minipage{1.0\textwidth}
\includegraphics[width=\textwidth,height=5cm]{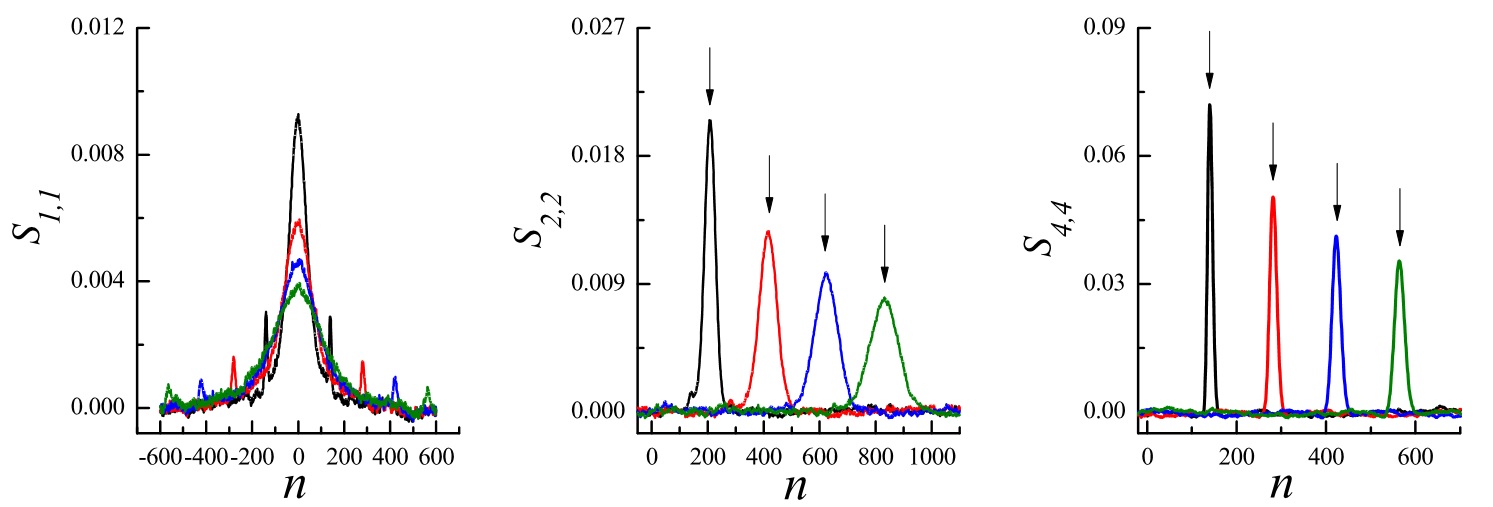} \endminipage
\vspace{0.2cm}
\minipage{1.0\textwidth}
\includegraphics[width=\textwidth,height=5cm]{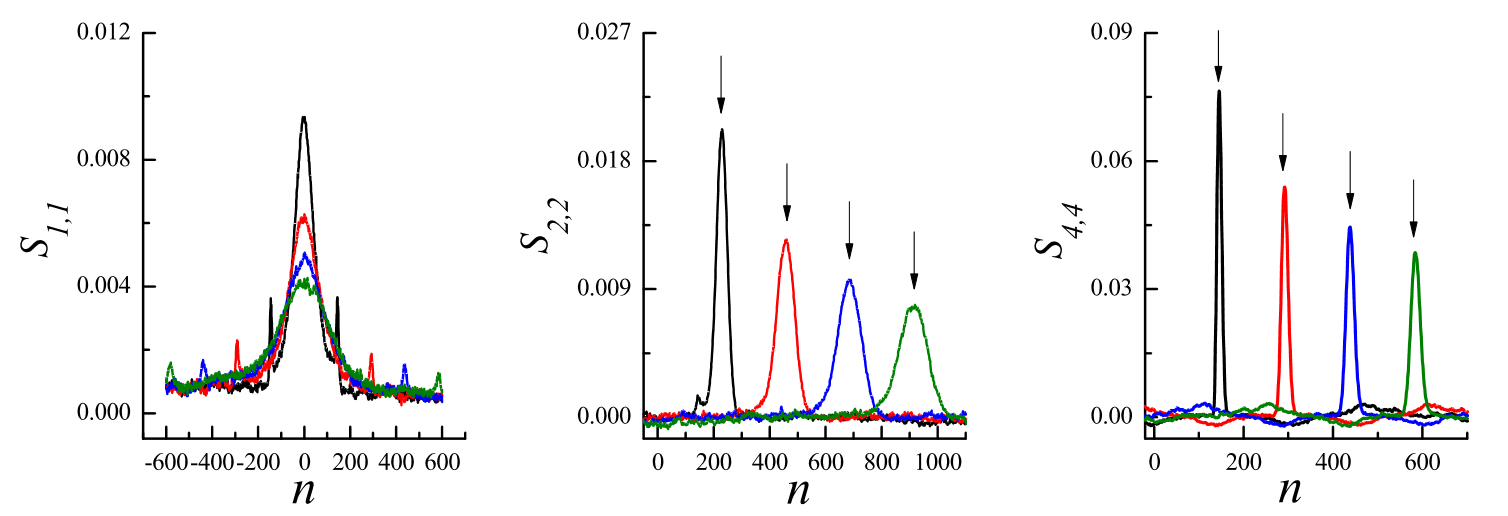} \endminipage
\caption{ (Color online) Autocorrelation functions at different times: Left panels ${\cal S}_{1,1}$ (heat mode), center panels ${\cal S}_{2,2}$ (longitudinal sound mode) and right panels ${\cal S}_{4, 4}$ (transverse sound mode). Vertical arrows indicate theoretical position for sounds peaks at $c t$. Temperature $T=1$, top/bottom panels correspond to elongation $\ell=1$/$\ell=1.1$.} 
\label{corrs}
\end{figure*}

We consider four sets of parameters that are representative of regimes for low and high temperatures with and without strain:\\
A) $T=0.1$ and $ \ell =1$ , \\
B) $T=0.1$ and $ \ell =1.1$ , \\
C) $T=1$ and $ \ell =1$ , \\
D) $T=1$ and $ \ell =1.1$ .

For each set we compute analytically the $R$ matrix in order to obtain the seven corresponding fields given by Eq. (\ref{matrizR}) as a function of time, from the elongation, momentum and energy.

To compute numerically the spatio-temporal correlations of these fields it is necessary to do a statistical average starting from a thermal equilibrated initial condition. For this purpose, we first integrate numerically the stochastic extension of Eqs. (\ref{EqMotionElongation}) and (\ref{EqMotionMomentum}) at the desired temperature, using a Runge-Kutta algorithm. After the equilibration was achieved, these equations are integrated in the microcanonical ensemble using a Velocity-Verlet algorithm. 

The last stage of the numerical integration allows to calculate the normal mode fields. Doing statistical average in time, we obtain the corresponding spatio-temporal correlations.

As the sound modes come in pairs travelling in opposite directions along the chain, and there is a degeneration in the transverse directions, we only show the correlations for the heat mode $\phi_1$, the longitudinal sound mode $\phi_2$ and the transverse sound mode $\phi_4$.

In Fig. \ref{corrs} we plot the mode correlations ${\cal S}_{1,1}$, ${\cal S}_{2,2}$ and ${\cal S}_{4,4}$ for the set of parameters C) and D) as a function of the site and for different times.

We observe that in all cases the peaks broaden and flatten for increasing times. The sound modes propagates along the chain with constant velocities while the heat mode remains at its initial position.

Also, for the heat peak, we observe the presence of two small lateral peaks, which corresponds to the coupling with the transverse sound mode due to the nonlinear terms in the hydrodynamic expansion. It is expected that at longer times these modes are decoupled.

From the position of the peaks, we compute numerically the velocities of the sound modes (see Table \ref{table_vel}), observing a good agreement with the theoretical predictions for all sets of parameters.

\begin{table}[h]
    \centering
    \begin{tabular}{||c||c|c||c|c||}
    \hline
        Set & $c_x$ Theo & $c_x$ Num& $c_{\perp}$ Theo &$ c_{\perp}$ Num\\
        \hline
          A & 1.12923 & 1.17 $\pm$ 0.01 & 0.608417 & 0.61 $\pm$ 0.01 \\
          B & 1.24087 & 1.37 $\pm$ 0.01 & 0.661516 & 0.66 $\pm$ 0.01 \\
          C & 2.05064 & 2.07 $\pm$ 0.02 & 1.40817 & 1.41 $\pm$ 0.01 \\
          D & 2.25594 & 2.30 $\pm$ 0.02 & 1.45634 & 1.46 $\pm$ 0.01 \\
         \hline
    \end{tabular}
    \caption{Theoretical and numerical velocities of longitudinal and transverse sound mode correlations.}
    \label{table_vel}
\end{table}

We obtain numerically the characteristic exponents $\delta_1$, $\delta_2$ and $\delta_4$ in Eqs. (\ref{exponentes}), from the broadening rate for each peak, which is the same as the flattening rate, within the numerical errors. The results are presented in Table \ref{tab_exp}.

\begin{table}[h]
    \centering
    \begin{tabular}{||c||c|c|c||}
    \hline
        Set & $\delta_1$ & $\delta_2$ & $\delta_4$\\
        \hline
          
          A) & 0.74 $\pm$ 0.02 & 0.70 $\pm$ 0.02 & 0.46 $\pm$ 0.01\\
          B) & 0.78 $\pm$ 0.02 & 0.70 $\pm$ 0.02 & 0.46 $\pm$ 0.01\\
          C) & 0.60 $\pm$ 0.02 & 0.69 $\pm$ 0.02 & 0.50 $\pm$ 0.01\\
          D) & 0.62 $\pm$ 0.02 & 0.69 $\pm$ 0.02 & 0.50 $\pm$ 0.01\\
         \hline
    \end{tabular}
    \caption{Scaling exponents for the heat, longitudinal and transverse sound modes.}
    \label{tab_exp}
\end{table}

\begin{figure*}
\centering
\includegraphics[width=0.9\textwidth,height=5cm]{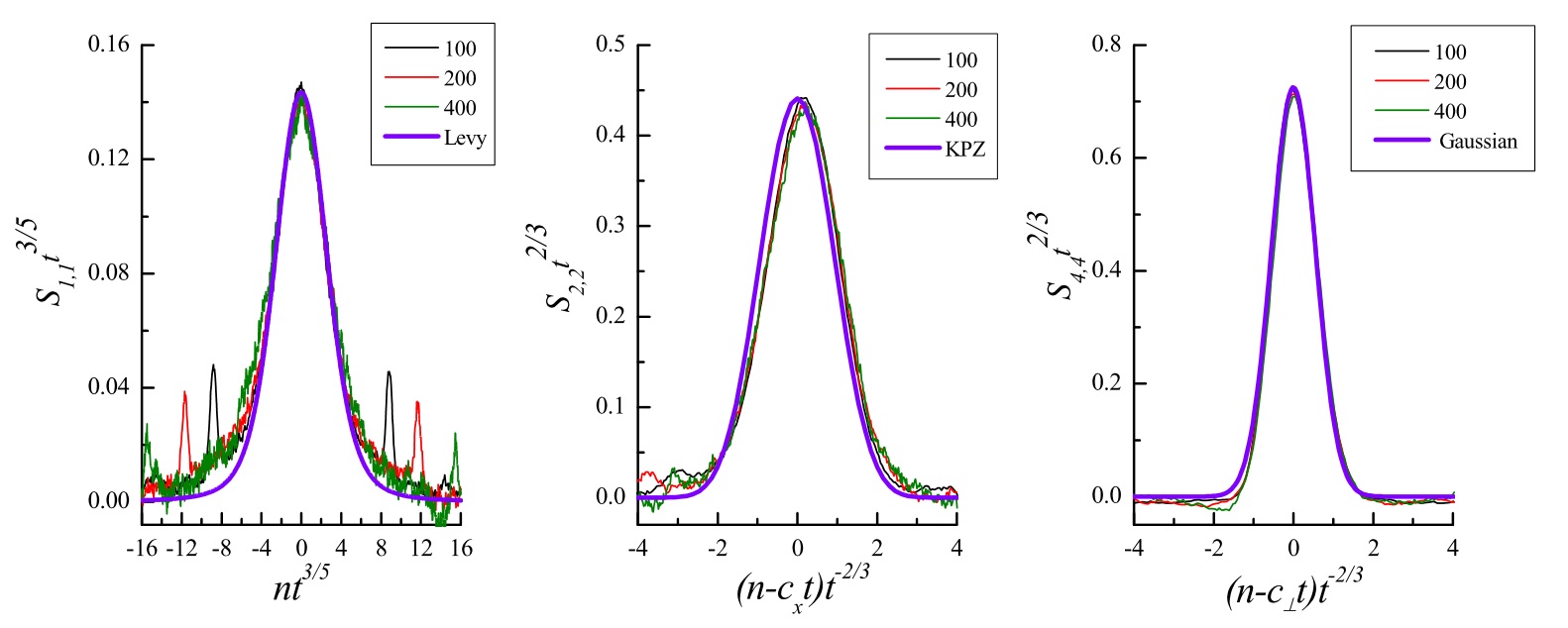}
\caption{(Color online) Rescaled and shifted autocorrelation functions for different times. Left: ${\cal S}_{1,1}$ (heat mode). A rescaled Levy $5/3$ function is superimposed.
Center: ${\cal S}_{2,2}$ (longitudinal sound mode). A rescaled KPZ function is superimposed. 
Right:  ${\cal S}_{4, 4}$ (transverse sound mode). A normalized Gaussian function with $\sigma=0.57$ is superimposed. $T=1$, $\ell=1$. } \label{heatcorrcolp}
\end{figure*}

We find for the heat mode and for the longitudinal sound mode that their exponents are in all cases larger that $1/2$ corresponding to a superdiffusive behavior. On the other hand for the transverse sound mode the exponent is close to $1/2$, corresponding to a normal diffusion.

For sets C and D, the numerical exponents agree with the theoretical predictions, while for sets A and B the exponents deviate, specially for the heat mode. These deviations at low temperatures are expected because the system explores regions around the minimum of the potential, where the non-linear terms are negligible.

On the other hand, at high temperatures the system is highly anharmonic, so NLFHT and MCT shoud work fine.

For the transverse sound mode, the normal diffusive behaviour gives a Gaussian scaling function, with an exponent $\delta_4 = 1/2$. 
For the longitudinal sound mode, as it was discussed previously, we expect a KPZ scaling function, with an exponent $\delta_2 = 2/3$, which is in near the numerical value.  

For the heat mode, we obtain characteristic exponents $\delta_1$ in the expected range between $3/5$ and $2/3$, but closer to $3/5$.

We show in Fig. \ref{heatcorrcolp}, for different times, the shifted and rescaled correlations peaks by their expected velocities and characteristic exponents. We also superimpose the expected corresponding scaling functions. It is observed a good agreement for the transverse and longitudinal sound modes peaks with a Gaussian and a KPZ function, respectively.
For the heat mode we propose a Levy function with an exponent $5/3$, which is closer to our numerical estimations. The deviations observed at the tails may indicate that the long time limit was not yet achieved. 

Numerically, we observe for high temperature, that the tension does not modify the characteristic exponents of the sound modes (within the statistical errors), nor their scaling functions (see Table \ref{tab_exp}). However, for the sound mode there is a slight dependence of its exponent with the tension.

\section{Conclusions}

In this paper we studied an anharmonic chain vibrating in 3 dimensions, with first neighbors interatomic potential that depends on the distance, and subjected to an external tension. This is a more realistic model that can be applied to e.g. suspended nanowires and polymers.

The 3D motion and the tension gives to our model a complex behavior in comparison with previous 1D anharmonic chain models. We applied the NLFHT where the new features of our model result in the emergence of two longitudinal and four transverse sound modes, added to a single heat mode. For longitudinal sound modes, the self-coupling term does not vanish, implying for its spatio-temporal correlation a superdiffusive behavior with a characteristic exponent of 2/3.

On the other hand, for a transverse sound mode, the self-coupling term vanishes, and it was necessary to apply MCT. Even within this theory, this mode does not couple to any others, resulting in a diffusive behavior for its spatio-temporal correlation.

Finally, for the heat mode, the self-coupling term also vanishes. MCT shows that this mode couples to all the sound modes, infering  a more complex scaling function with characteristic exponent in between 3/5 and 2/3.

As it is expected, we have checked numerically that this theory does not work at low temperatures, where the particles feels an almost harmonic potential.

For high temperatures, the numerical simulations are in a reasonable good agreement with the theoretical predictions for the sound velocities and the exponents. For the heat mode the exponent is closer to 3/5, but it seems to slightly depend on the tension, which affects the couplings with the sound modes. The role of the tension, as well as the boundary conditions, deserve further investigation.

These results for the correlations of the fields can be applied to obtain the thermal conductivity of the chain through the Green-Kubo formula, suggesting an anomalous thermal conduction.

\section*{Acknowledgments}
The authors want to thank for the funding project PIO-CONICET 14420140100013CO.

\section*{Appendix}
\setcounter{equation}{0}
\renewcommand{\theequation}{A\arabic{equation}}

For a general derivative with respect to the microcanonical variables $\ell$ and $e$ of a quantity $h$, we have

\begin{eqnarray}
\frac{\partial h}{\partial \ell} &=& \frac{\partial \beta}{\partial \ell} \frac{\partial h}{\partial \beta} + \frac{\partial F}{\partial \ell} \frac{\partial h}{\partial F} \ , \\
\frac{\partial h}{\partial e} &=& \frac{\partial \beta}{\partial e} \frac{\partial h}{\partial \beta} + \frac{\partial F}{\partial e} \frac{\partial h}{\partial F} \ .
\end{eqnarray}
Additionally, if $h$ is an average on the canonical ensemble, we have the following rules for derivation with respect to $\beta$ and $F$

\begin{eqnarray}
\frac{\partial \langle h \rangle}{\partial \beta} &=& F ( \langle h x \rangle - \langle h \rangle \langle x \rangle ) - ( \langle h V \rangle - \langle h \rangle \langle V \rangle ) \nonumber \\
&=& F \langle h \ \delta x \rangle - \langle h \ \delta V \rangle \ , \\
\frac{\partial \langle h \rangle}{\partial F} &=& \beta ( \langle h x \rangle - \langle h \rangle \langle x \rangle ) = \beta \langle h \ \delta x \rangle \ .
\end{eqnarray}
These rules allow to compute the derivatives of the external force, needed for the $A$ matrix, which are obtained from the following expressions:
\begin{eqnarray}
\frac{\partial \ell}{\partial F} &=& \frac{\partial \langle x \rangle}{\partial F} = \beta C_{1 1} \ , \nonumber \\
\frac{\partial e}{\partial F} &=& \frac{\partial \langle \epsilon \rangle}{\partial F} = \beta C_{1 7} \ , \nonumber \\
\frac{\partial \ell}{\partial \beta} &=& \frac{\partial \langle x \rangle}{\partial \beta} = F C_{1 1} - C_{1 7} \ , \nonumber \\
\frac{\partial e}{\partial \beta} &=& \frac{\partial \langle \epsilon \rangle}{\partial \beta} = F C_{1 7} - C_{7 7} \ . \label{PartialsLE}
\end{eqnarray}
Using that $F = F ( \ell(\beta, F) , e(\beta, F) )$ and $\beta = \beta ( \ell(\beta, F) , e(\beta, F) )$ it is possible to determine that
\begin{eqnarray}
\frac{\partial F}{\partial \ell} &=& - \frac{1}{\Gamma} \frac{\partial e}{\partial \beta} \ , \nonumber \\
\frac{\partial F}{\partial e} &=& \frac{1}{\Gamma} \frac{\partial \ell}{\partial \beta} \ , \nonumber \\
\frac{\partial \beta}{\partial \ell} &=& \frac{1}{\Gamma} \frac{\partial e}{\partial F} \ , \nonumber \\
\frac{\partial \beta}{\partial e} &=& -\frac{1}{\Gamma} \frac{\partial \ell}{\partial F} \ ,
\end{eqnarray} 
with
\begin{equation}
\Gamma = \frac{\partial e}{\partial F} \frac{\partial \ell}{\partial \beta} - \frac{\partial e}{\partial \beta} \frac{\partial \ell}{\partial F} \ .
\end{equation}
Replacing Eqs. (\ref{PartialsLE}) we obtain
\begin{eqnarray}
\frac{\partial F}{\partial \ell} &=& \frac{ C_{7 7} - F C_{1 7} }{ \beta ( C_{1 1} C_{7 7} - C_{1 7}^2) } \ , \nonumber \\
\frac{\partial F}{\partial e} &=& \frac{ F C_{1 1} - C_{1 7} }{ \beta ( C_{1 1} C_{7 7} - C_{1 7}^2) } \ , \nonumber \\
\frac{\partial \beta}{\partial \ell} &=& \frac{ C_{1 7} }{ C_{1 1} C_{7 7} - C_{1 7}^2 } \ , \nonumber \\
\frac{\partial \beta}{\partial e} &=& \frac{ - C_{1 1} }{ C_{1 1} C_{7 7} - C_{1 7}^2 } \ .
\end{eqnarray}

We can also compute the derivatives of the correlators of order two
\begin{eqnarray}
\frac{\partial C_{1 1} }{\partial \beta} &=& F \langle \delta x^3 \rangle - \langle \delta x^2 \ \delta V \rangle \ , \nonumber \\
\frac{\partial C_{1 1} }{\partial F} &=& \beta \langle \delta x^3 \rangle \ , \nonumber \\
\frac{\partial C_{1 7} }{\partial \beta} &=& F \langle \delta x^2 \ \delta V \rangle - \langle \delta x \ \delta V^2 \rangle \ , \nonumber \\
\frac{\partial C_{1 7} }{\partial F} &=& \beta \langle \delta x^2 \ \delta V \rangle \ , \nonumber \\
\frac{\partial C_{7 7} }{\partial \beta} &=& - \frac{3}{\beta^3} + F \langle \delta x \ \delta V^2 \rangle - \langle \delta V^3 \rangle \ , \nonumber \\
\frac{\partial C_{7 7} }{\partial F} &=& \beta \langle \delta x \ \delta V^2 \rangle \ , 
\end{eqnarray}
in terms of correlators of order three. These results allow to compute the second derivatives of the external force.


For the $A$ matrix, the right eigenvectors are
\begin{eqnarray}
\vec{v}_{\textnormal{R} 1} &=& \frac{1}{Z_1} \left( F_e, 0, 0, 0, 0, 0, - F_{\ell} \right) \ , \nonumber  \\
\vec{v}_{\textnormal{R} 2} &=& \frac{1}{Z_2} \left( 1, 0, 0, -m c_x, 0, 0, F \right) \ , \nonumber \\
\vec{v}_{\textnormal{R} 3} &=& \frac{1}{Z_3} \left( 1, 0, 0, +m c_x, 0, 0, F \right) \ , \nonumber \\
\vec{v}_{\textnormal{R} 4} &=& \frac{1}{Z_4} \left( 0, 1, 0, 0, -m c_{\perp}, 0, 0 \right) \ , \nonumber \\
\vec{v}_{\textnormal{R} 5} &=& \frac{1}{Z_5} \left( 0, 1, 0, 0, +m c_{\perp}, 0, 0 \right) \ , \nonumber \\
\vec{v}_{\textnormal{R} 6} &=& \frac{1}{Z_6} \left(0, 0, 1, 0, 0, -m c_{\perp}, 0 \right) \ , \nonumber \\
\vec{v}_{\textnormal{R} 7} &=& \frac{1}{Z_7} \left(0, 0, 1, 0, 0, +m c_{\perp}, 0 \right) \ ,
\end{eqnarray}
with $Z_i$ normalization constants. As columns, they form the matrix $R^{-1}$. The left eigenvectors of $A$ are
\begin{eqnarray}
\vec{v}_{\textnormal{L} 1} &=& \frac{Z_1}{m c_x^2} \left( F, 0, 0, 0, 0, 0, - 1 \right) \ , \nonumber  \\
\vec{v}_{\textnormal{L} 2} &=& \frac{Z_2}{2 m c_x^2} \left(F_{\ell}, 0, 0, - c_x, 0, 0, F_e \right) \ , \nonumber \\
\vec{v}_{\textnormal{L} 3} &=& \frac{Z_3}{2 m c_x^2} \left(F_{\ell}, 0, 0, + c_x, 0, 0, F_e \right) \ , \nonumber \\
\vec{v}_{\textnormal{L} 4} &=& \frac{Z_4}{2} \left( 0, 1, 0, 0, -\frac{1}{m c_{\perp}}, 0, 0 \right) \ , \nonumber \\
\vec{v}_{\textnormal{L} 5} &=& \frac{Z_5}{2} \left( 0, 1, 0, 0, +\frac{1}{m c_{\perp}}, 0, 0 \right) \ , \nonumber \\
\vec{v}_{\textnormal{L} 6} &=& \frac{Z_6}{2} \left( 0, 0, 1, 0, 0, -\frac{1}{m c_{\perp}}, 0 \right) \ , \nonumber \\
\vec{v}_{\textnormal{L} 7} &=& \frac{Z_7}{2} \left( 0, 0, 1, 0, 0, +\frac{1}{m c_{\perp}}, 0 \right) \ ,
\end{eqnarray}
which as rows, they form the matrix $R$. The constants $Z_i$ can be determined uniquely (up to a sign) imposing that the correlation matrix in the mode basis is the identity $(R C R^{\textnormal{T}} = I)$
\begin{eqnarray}
Z_1^2 &=& \frac{F^2 C_{1 1} - 2 F C_{1 7} + C_{7 7} }{ \beta^2 ( C_{1 1} C_{7 7} - C_{1 7}^2)^2 } = \frac{m c_x^2}{\Gamma} \ , \\
Z_2^2 &=&  Z_3^2 = \frac{2 (F^2 C_{1 1} - 2 F C_{1 7} + C_{7 7}) }{  C_{1 1} C_{7 7} - C_{1 7}^2 } \\
&=& 2 m \beta c_x^2 \nonumber \ , \\
Z_4^2 &=&  Z_5^2 = Z_6^2 =  Z_7^2 = \frac{2 \beta F}{\ell} = 2 m \beta c_{\perp}^2 \ .
\end{eqnarray}

The Hessian matrices are explicitly
\begin{equation}
H^1_{\alpha \beta} = H^2_{\alpha \beta} = H^3_{\alpha \beta} = 0 \ ,
\end{equation}

\begin{equation}
H^4_{\alpha \beta} = \left(
\begin{array}{ccccccc}
-F_{\ell \ell } & 0 & 0 & 0 & 0 & 0 & -F_{e \ell} \\
0 & \frac{F}{\ell^2} - \frac{F_{\ell}}{\ell} & 0 & 0 & 0 & 0 & 0 \\
0 & 0 & \frac{F}{\ell^2} - \frac{F_{\ell}}{\ell} & 0 & 0 & 0 & 0 \\
0 & 0 & 0 & \frac{F_e}{m} & 0 & 0 & 0 \\
0 & 0 & 0 & 0 & \frac{F_e}{m} & 0 & 0 \\
0 & 0 & 0 & 0 & 0 & \frac{F_e}{m} & 0 \\
-F_{e \ell} & 0 & 0 & 0 & 0 & 0 & -F_{e e}
\end{array} 
\right) \ ,
\end{equation}

\begin{equation}
H^5_{\alpha \beta} = \left(
\begin{array}{ccccccc}
0 & \frac{F}{\ell^2} - \frac{F_{\ell}}{\ell}  & 0 & 0 & 0 & 0 & 0 \\
\frac{F}{\ell^2} - \frac{F_{\ell}}{\ell}  & 0 & 0 & 0 & 0 & 0 & -\frac{F_e}{\ell} \\
0 & 0 &0 & 0 & 0 & 0 & 0 \\
0 & 0 & 0 & 0 & 0 & 0 & 0 \\
0 & 0 & 0 & 0 & 0 & 0 & 0 \\
0 & 0 & 0 & 0 & 0 &0 & 0 \\
0 & -\frac{F_e}{\ell} & 0 & 0 & 0 & 0 & 0
\end{array} 
\right) \ ,
\end{equation}

\begin{equation}
H^6_{\alpha \beta} = \left(
\begin{array}{ccccccc}
0 & 0 & \frac{F}{\ell^2} - \frac{F_{\ell}}{\ell}   & 0 & 0 & 0 & 0 \\
0 & 0 &0 & 0 & 0 & 0 & 0 \\
\frac{F}{\ell^2} - \frac{F_{\ell}}{\ell}  & 0 & 0 & 0 & 0 & 0 & -\frac{F_e}{\ell} \\
0 & 0 & 0 & 0 & 0 & 0 & 0 \\
0 & 0 & 0 & 0 & 0 & 0 & 0 \\
0 & 0 & 0 & 0 & 0 &0 & 0 \\
0 & 0& -\frac{F_e}{\ell}  & 0 & 0 & 0 & 0
\end{array} 
\right)\ ,
\end{equation}

\begin{equation}
H^7_{\alpha \beta} = \left(
\begin{array}{ccccccc}
0 & 0 & 0 & - \frac{F_{\ell}}{m} & 0 & 0 & 0 \\
0 & 0 & 0 & 0 & -\frac{F}{m l} & 0 & 0 \\
0 & 0 & 0 & 0 & 0 & -\frac{F}{m l} & 0 \\
- \frac{F_{\ell}}{m} & 0 & 0 & 0 & 0 & 0 & - \frac{F_e}{m} \\
0 & -\frac{F}{m l} & 0 & 0 & 0 & 0 & 0 \\
0 & 0 & -\frac{F}{m l} & 0 & 0 & 0 & 0 \\
0 & 0 & 0 & - \frac{F_e}{m}  & 0 & 0 & 0
\end{array} 
\right) \ .
\end{equation}

The diffusion matrix has the following structure:

\begin{equation} \label{DiffusionMatrix}
D_{\alpha \beta} = 
\left(
\begin{array}{ccccccc}
D_{1 1} & 0 & 0 & 0 & 0 & 0 & D_{1 7} \\
0 & 0 & 0 & 0 & 0 & 0 & 0 \\
0 & 0 & 0 & 0 & 0 & 0 & 0 \\
0 & 0 & 0 & \frac{\beta \sigma_{p_x}^2 }{2 m} & 0 & 0 & 0 \\
0 & 0 & 0 & 0 & \frac{\beta \sigma_{p_{\perp}}^2 }{2 m} & 0 & 0 \\
0 & 0 & 0 & 0 & 0 & \frac{\beta \sigma_{p_{\perp}}^2 }{2 m} & 0 \\
D_{1 7} & 0 & 0 & 0 & 0 & 0 & D_{7 7}
\end{array} 
\right) \ ,
\end{equation}
with
\begin{eqnarray}
D_{1 1} &=& \frac{ C_{1 7}^2 \sigma_{e}^2 }{2 (C_{1 1} C_{7 7} - C_{1 7}^2) (C_{1 1} + C_{7 7})} \ ,\\
D_{1 7} &=& - \frac{ C_{1 1} C_{1 7}  \sigma_{e}^2 }{2 (C_{1 1} C_{7 7} - C_{1 7}^2) (C_{1 1} + C_{7 7})} \ ,\\
D_{7 7} &=& \frac{ ( C_{1 1} C_{7 7} - C_{1 7}^2 + C_{1 1}^2 ) \sigma_{e}^2 }{2 (C_{1 1} C_{7 7} - C_{1 7}^2) (C_{1 1} + C_{7 7})}\ .
\end{eqnarray}

\end{document}